\documentclass[fleqn,usenatbib]{mnras}

\usepackage{newtxtext,newtxmath}

\usepackage[T1]{fontenc}

\DeclareRobustCommand{\VAN}[3]{#2}
\let\VANthebibliography\thebibliography
\def\thebibliography{\DeclareRobustCommand{\VAN}[3]{##3}\VANthebibliography}

\usepackage{graphicx}	
\usepackage{amsmath}	

\title[Variables in the bulge cluster NGC 6558]{The variable stars in the field of the bulge cluster NGC 6558}

\author[Arellano Ferro et al.]{
A. Arellano Ferro,$^{1}$\thanks{Corresponding Author: E-mail: armando@astro.unam.mx}
L. J. Zerpa Guillen,$^{2,3}$
M. A. Yepez,$^{1,4}$ 
I. H. Bustos Fierro,$^{5}$ 
\and
Z. Prudil,$^{6}$
C. E. P\'erez Parra,$^{2,3}$
\\
$^{1}$Instituto de Astronom\'ia, Universidad Nacional Aut\'onoma de M\'exico, Ciudad de M\'exico, CP 04510, M\'exico.\\
$^{2}$Universidad de Los Andes, Facultad de Ciencias, Dpto. Física, Grupo de Astrofísica Teórica, Mérida,
Venezuela.\\
$^{3}$Fundaci\'on Centro de Investigaciones de Astronomía Francisco, J. Duarte (CIDA), Mérida, Venezuela.\\
$^{4}$Instituto Nacional de Astrofísica, Óptica y Electrónica (INAOE), Luis Enrique Erro No.1, Tonantzintla, Pue., C.P. 72840, México\\
$^5$ Observatorio
Astron\'omico, Universidad Nacional de Córdoba, Córdoba C.P. 5000, Argentina.\\
$^{6}$ European Southern Observatory, , Karl-Schwarzschild-Straße 2, 85748, Garching, Germany\\
}

\date{Accepted: ; Received: 2020  in original form 2020 July}

\pubyear{2020}

\begin{document}

\label{firstpage}
\pagerange{\pageref{firstpage}--\pageref{lastpage}}
\maketitle

\begin{abstract}
We made a survey of the variable stars in a 13.2 x 13.2 arcmin$^{2}$ centered on the field of the Galactic bulge cluster NGC 6558. A total of 78 variables was found in the field of the cluster. Many of these variables are included in the Catalogue of Variable Stars in Galactic Globular Clusters (Clement et al. 2001), OGLE or G$aia$-DR3 data releases.  A membership analysis based on the proper motions of G$aia$-DR3 revealed that many of these variables do not belong to the cluster. We employed the data from the aforementioned surveys and our own data in the \emph{VI} photometric system to estimate the periods, which along with the light curves morphology and position in a deferentially dereddened colour- magnitude diagram(CMD), help classifying the variable types. Two new member variables were found; an eclipsing binary (V18) and a semi-regular SR/L (V19). In the end we conclude that only 9 variables are likely cluster members.  Member variables were used to discuss the mean metallicity and distance of the parental cluster and find the average values .
\end{abstract}



\maketitle

\section{Introduction}
\label{sec:intro}

The globular cluster NGC 6558 resides in the bulge of the Galaxy, very near to the Galactic center \citep{Bica2016}. From high resolution spectroscopic analysis of four red giants in NGC 6558 \citet{Barbuy2018} concluded that the cluster shows  abundance pattern typical of the oldest inner bulge clusters and could be among the oldest objects in the Galaxy. Dynamically, the orbit of NGC 6558 is confined to less that 1.5 kpc (see the orbital integration in Fundamental parameters of Galactic globular cluster by 
Baumgardt et al. (2023)\footnote{%
 \texttt{https://people.smp.uq.edu.au/HolgerBaumgardt/globular/}}from the Galactic center, confirming the cluster is a local resident with all  rights.

Being the bulge regions very rich in stars and dust, the images of bulge clusters are highly contaminated with field stars and are generally subject to heavy interstellar differential reddening. Hence, to properly study the colour magnitude Diagram (CMD) of a bulge cluster, its stellar populations or families of variable stars, or to use specific groups of stars as indicators of physical properties, requires a thorough membership analysis and the calculation of a reddening map across the field of the system (e.g.  \citet{AlonsoGarcia2012}; \citet{Yepez2023}). Once this is achieved, then some of the variable stars can be used as indicators of cluster metallicity and distance, and their positions in the CMD can be compared with theoretical calculations to inferred evolutionary stages and internal stellar structure, in particular for the stars in the horizontal branch (HB). Previous studies carried out by  our team towards the estimation of the main physical parameters; reddening, distance and metallicity employing their variable star populations are described in detail in the works by \citet{Arellano2022} and \citet{Arellano2024}. The results for the bulge clusters  NGC 6333, NGC 6401 and NGC 6522 are reported in the papers by \cite{Arellano2013}, \cite{Tsapras2017} and \cite{Arellano2023} respectively.

\section{Observations and data reduction}
\label{sec:Obser}

\subsection{Bosque Alegre and Las Campanas Data}

The data for this work were obtained with the 1.54 m telescope of the Bosque Alegre Astronomical Station (EABA), of the National Observatory of Cordoba, Argentina, during August-September 2018 and  between June and August 2019, for a total of eleven nights. Two detectors were employed; in 2018 a CCD KAF-16803 with 4096 × 4096 pixeles, while in 2019 a CCD KAF-6303E with 3072 × 2048 pixeles.The corresponding fields were 16.9 × 16.9 arcmin$^2$ and 12.6 × 8.4 arcmin$^2$ respectively. Also observations were performed with the 1m telescope SWOPE from Las Campanas Observatory, Chile, during three nights in June 2018. The dectector was a CCD E2V 231-81 with 4096 x 4112 pixels and a field of 14,8 × 14,9 arcmin$^2$. See Table \ref{log} for the log of the observations. In the following we shall refer to EABA seasons as BA18 and BA19, and to Las Campanas seasons as SWOPE.

\begin{table}
\footnotesize
\caption {Log of the observation of NGC 6558$^*$.}
\label{log}
\centering
\begin{tabular}{ccccccc}
\hline
Date & Site	& $N_V$ &$t_V (s)$ &$N_I$ & $t_I (s)$ &Avg.\\
   &   &          & sec    &         &sec&seeing (")\\
\hline
2018-06-27	& SWOPE	& 143 & 4 - 60 & 138 & 2 - 10 & 1.05 \\
2018-06-29	& SWOPE & 101 & 4 - 80 & 98  & 2 - 10 & 1.37  \\
2018-06-30	& SWOPE & 47 & 10 - 80 & 41  & 5 - 10 & 1.48  \\
2018-08-04	& EABA & 44 & 120 - 200 & 39  & 100 - 60 & 2.65  \\
2018-08-06	& EABA & 48 & 120 & 51  & 60 & 2.61  \\
2018-08-12	& EABA & 39 & 120 & 39  & 60 & 3.38  \\
2018-09-03	& EABA & 32 & 120 & 35  & 60 & 2.31  \\
2018-09-15	& EABA & 21 & 120 & 27  & 60 & 2.38  \\
2018-09-17	& EABA & 33 & 120 & 33  & 60 & 2.70  \\
2019-06-29	& EABA & 43 & 120 & 53  & 60 & 3.19  \\
2019-07-27	& EABA & 28 & 120 & 33  & 60 & 2.22  \\
2019-07-28	& EABA & 54 & 120 & 51  & 60 & 2.44  \\
2019-08-03	& EABA & 34 & 120 & 41  & 60 & 2.39  \\
2019-08-10	& EABA & 51 & 120 & 52  & 60 & 2.95  \\
\hline
Total:&  & 718 &  & 731  &  &   \\
\hline
\end{tabular}
\raggedright
\quad $*$
Columns $N_V$ and $N_I$ record the number of images acquired while $t_v$ y $t_I$ indicate the typical exposure times. The average nightly seeing is given in the last column.
\end{table}

\subsection{Data Reduction}
\label{DIA}

To extract high-precision time-series photometry for all point sources in the field of our images, we employed the Difference Image Analisis (DIA) and the pipeline DanDIA \citep{Bramich2008}; \citep{Bramich2013}; \citep{Bramich2015}.. 
The reference $V$ and $I$ images are obtained by stacking up images of the best quality in the collection. Then, individual images are subtracted from the reference to create individual differential images where the flux
of all detected point sources is measured. The total flux in ADU/s is calculated as:

\begin{equation}
f_{\rm tot} (t) = f_{\rm ref} + \frac{f_{\rm diff} (t)} {p (t)},
\end{equation}

\noindent
where $f_{\rm ref}$ is the reference flux (ADU/s), $f_{\rm diff} (t)$ the differential flux (ADU/s) and $p(t)$ is the
the photometric scale factor. To convert fluxes to instrumental magnitudes we used: 

\begin{equation}
m_{\rm ins} (t) = 25.0 - 2.5 log[f_{\rm tot}(t)],
\end{equation}

\noindent
where $m_{ins}(t)$ is the instrumental magnitude of the star at time $t$.

\subsection{OGLE and $Gaia$-DR3 photometric data}

We have also extensively used data of the variable stars in the field of the cluster identified in the OGLE and $Gaia$-DR3 data bases, and shall be discussed in the following sections.

\subsection{Transformation to the \textit{VI} standard system}

The instrumental light curves can be converted into the standard system employing local standard stars in the field of the cluster. We found 9, 10 and 12 standard stars for the images in the seasons SWOPE, EABA2018 and EABA2019 respectivelly, in the catalogue of \citep{Stetson2000}\footnote{%
\texttt{https://www.canfar.net/storage/list/STETSON/Standards}}, which have been set into the Johnson-Kron-Cousins standard system using the equatorial standards from  \citet{Landolt1992}. The transformation equations are of the form $V$ – $v$ = A($v$ – $i$) + B, and  $I$ – $i$ = C($v$ – $i$) + D, and the season constants are reported in Table \ref{tab2}.

\begin{table} 
\scriptsize
\begin{center}
\caption{Season coefficients in transformation equation of the form  $V$–$v$ = A($v$-$i$) + B, and $I$ – $i$ = C($v$ – $i$) + D.}
\label{tab2}

\begin{tabular}{cccc}
\hline
Coeff.& SWOPE	& EABA 2018 & EABA 2019 \\
\hline
A	& $-1.724 \pm 0.0194$ & $-2.726 \pm 0.093$ & -3.19804±0.04664 \\
B	& $-0.101 \pm 0.016$ & $0.113 \pm 0.139$ &  $0.15109 \pm 0.05390$ \\
C 	& $-4.172 \pm 0.855$ & $-3.589 \pm 0.101$ &  -3.94014±0.04784 \\
\hline
\end{tabular}
\end{center}
\end{table}

In Table \ref{tab:vi_phot} we include a small portion of the time-series \emph{VI} photometry obtained in this work. The full table shall be available in electronic form in the Centre de Donnés astronomiques de Strasbourg database (CDS).

\begin{table}
\scriptsize
\begin{center}
\caption{Time-series \textit{VI} photometry for the variables stars observed in this work$^*$}
\label{tab:vi_phot}
\centering
\begin{tabular}{cccccc}
\hline
Variable &Filter & HJD & $M_{\mbox{\scriptsize std}}$ &
$m_{\mbox{\scriptsize ins}}$
& $\sigma_{m}$ \\
Star ID  &    & (d) & (mag)     & (mag)   & (mag) \\
\hline
 V1 & $V$& 2458342.52058& 11.849 & 14.402 & 0.001 \\   
 V1 & $V$& 2458342.52211& 11.849 & 14.402 & 0.001 \\
\vdots   &  \vdots  & \vdots & \vdots & \vdots & \vdots  \\
 V1 & $I$ & 2458342.51828 & 10.260 & 13.681 & 0.001\\  
 V1 & $I$ & 2458342.51870 & 10.256 & 13.677 & 0.001  \\ 
\vdots   &  \vdots  & \vdots & \vdots & \vdots & \vdots  \\
 V3 & $V$ & 2458727.54934 & 12.591& 15.755 & 0.002 \\   
 V3 & $V$ & 2458727.55030 & 12.591& 15.755&  0.001 \\
\vdots   &  \vdots  & \vdots & \vdots & \vdots & \vdots  \\
 V3 & $I$ & 2458727.54759 & 11.381&  14.984 & 0.002 \\    
 V3 & $I$ & 2458727.54809 & 11.387&  14.991 & 0.001 \\   
\vdots   &  \vdots  & \vdots & \vdots & \vdots & \vdots  \\
\hline
\end{tabular}
\end{center}
* The standard and
instrumental magnitudes are listed in columns 4 and~5,
respectively, corresponding to the variable stars in column~1. Filter and epoch of
mid-exposure are listed in columns 2 and 3, respectively. The uncertainty on
$\mathrm{m}_\mathrm{ins}$, which also corresponds to the
uncertainty on $\mathrm{M}_\mathrm{std}$, is listed in column~6. A full version of this table is available at the CDS database.

\end{table}

\section{Membership anaysis}
\label{membership}

Being in the Galactic bulge, the colour-magnitude diagram (CMD) of NGC 6558 is heavily contaminated by field stars all   subject to a remarkable differential reddening. To produce a clean and useful CMD a membership analysis and a local reddening map are in order.

\begin{figure*}
\includegraphics[width=17.5 cm]{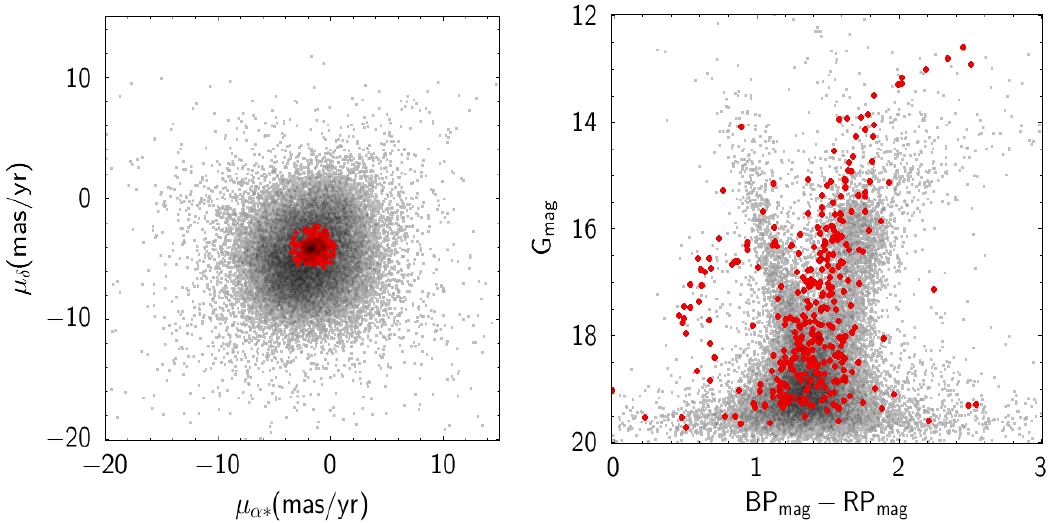}
\caption{Results from the membership analysis (section \ref{membership}). Left and right panels show the vector point (VPD) and CMD diagrams respectively. Red dots represent the likely cluster members while the gray dots represent the field stars.} 
\label{mem_analysis}
\end{figure*}

The membership analysis was performed using the positions and proper motions available in the $Gaia$-DR3 and employing the method of \citet{Bustos2019}. The method is based on a two step approach: 1) it finds groups of stars with similar characteristics in the four-dimensional space of the gnomonic coordinates ($X_{\rm t}$,$Y_{\rm t}$) and proper motions ($\mu_{\alpha*}$,$\mu_\delta$) employing the BIRCH clustering algorithm \citep{Zhang1996} and 2) in order to extract likely members that were missed in the first stage, the analysis of the projected distribution of stars with different proper motions around the mean proper motion of the cluster is performed.

There are 45,506. point sources within an 8 arcmin field centered in the cluster. However, proper motions are available for only 28,334. Most of the stars were found field stars and only 495 were identified as likely cluster members. Fig. \ref{mem_analysis} displays the results in the Vector-Point diagram (VPD) and the resulting Colour-Magnitude diagram (CMD). The identified likely members clearly trace the major features in the CMD, such as the horizontal branch (HB) and the red giant branch (RGB).

\section{Differential reddening and the CMD }
\label{differential}

To properly deredden the colour magnitude diagram of this bulge cluster, it is necessary to consider the effects of differential reddening. Fortunately, NGC 6558 has been included in the thorough reddening study of the inner Galaxy by \citet{AlonsoGarcia2012}. We used their reddening map to differentially deredden the \emph{VI} CMD obtained from our photometry. 

\begin{figure*}
\includegraphics[width=17.5 cm]{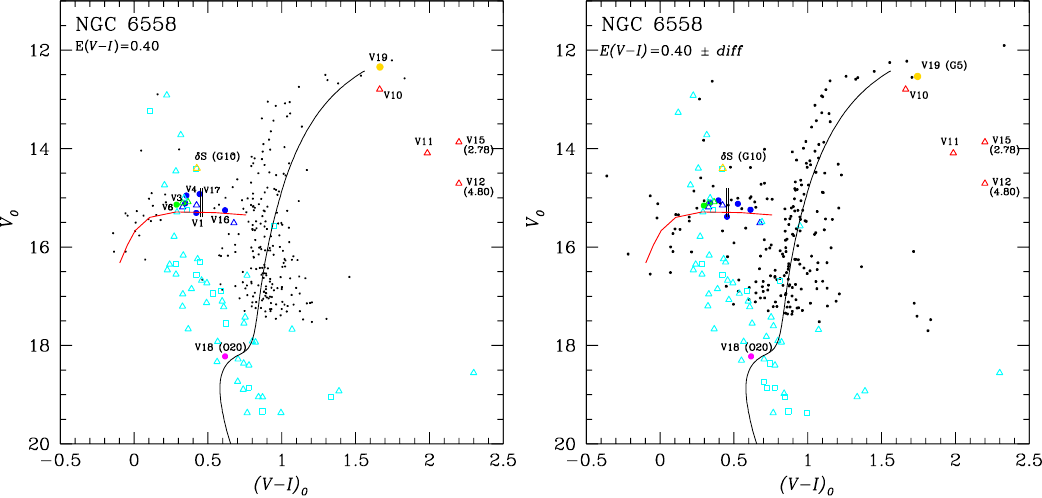}
\caption{The CMDs built from cluster members measured by our \emph{VI} photometry. The panel on the left shows the CMD dereddened with an average $E(B-V)=0.40$ while the one to the right has been differentially dereddened as described in section \ref{differential}. The improvement after the corrections is clear. Variable stars in the field of the cluster, members and no members, are shown by colour symbols according to the following code: RRab-blue; RRc-green; RGB-red; $\delta$ Sct (G10) and member RGB (V19) -yellow; binaries -turquoise; member binary V18 (O20)-magenta. Solid symbols are used for members and empty ones for field stars (triangles) or stars of unknown status due to lack of proper motion (squares). The isochrone is from \citet{Vandenberg2014} for [Fe/H]=$-1.35$ and an age of 12.0 Gyrs. Red ZAHB is from the models built from the Eggleton code \citep{Pols1997,Pols1998, KPS1997}, and calculated by \citet{Yepez2022}. The vertical black lines at the ZAHB mark the empirical red edge of the first overtone instability strip  \citep{Arellano2015,Arellano2016}. } 
\label{CMDs}
\end{figure*}  

In Fig. \ref{CMDs} we display on the left panel  the CMD dereddened with a constant value $E(B-V)=0.40$. On the right panel we have applied the differential corrections guided by the reddening map, i.e. $E(B-V) + diff$, where $diff$ corresponds to the differential corrections for the corresponding position of the star. The improvement on the  dispersion at the RGB and the HB are obvious

\section{The variable stars in NGC 6558}

The field of NGC 6558, like in most of the bulge clusters, is very rich in variable stars. The advent of missions like $Gaia$ and OGLE have detected a large number of them. However, many of these variables are not linked to the cluster but are merely projected against it. Our aim in the following sections is to critically evaluate the membership of them. The challenge is to achieve good photometric values that entitle us to position the variables in our observed and properly dereddened \emph{VI} CMD, which along with the proper motion analysis, their pulsational type and, in the case of RR Lyrae, the estimation of their distance via the Fourier decomposition of their light curves, should allow us to pronounce about their membership status.

\subsection{The catalogued variables (Clement et al. 2001)}

The catalogue of variable stars in globular clusters (CVSGC) \citep{Clement2001} in its 
April 2016 update, lists 17 variables and labels two of them (V2 and V7) as probably non-variable. The rest are 8 RRab, 3 RRc, and 4 long period variables labeled "L?"

\subsection{Other variables in the field of NGC 6558 according to OGLE, and $Gaia$}

In the $Gaia$ mission \citep{Gaia2016} and the Optical Gravitational Lensing Experiment (OGLE) \citep{Udalski1992} many variables were detected in the field of NGC 6558. We have cross-matched those variables with point sources measured in our photometry and have combined the photometric data to build the light curves. In the case of OGLE III \citep{Soszynski2013} and IV \citep{Soszynski2014}, \emph{VI} magnitudes are available. For the case of $Gaia$-DR3 data, a transformation into the \emph{VI} system was necessary. This was achieved using the transformation equations of \citep{Riello2021}.

We identified 56 variables in the OGLE data base within the field of NGC 6558 and for the present purpose of the paper we shall identify them with the prefix 'O'. Similarly we noticed 13 stars announced as variable in $Gaia$-DR3, these stars we call them with the prefix 'G'. Then we aim to confirm their variability and type, and to check their membership. The table cross matching the variables in the CVSGC, OGLE and $Gaia$, indicating their membership status, and their light curves  are reported in Appendix A at the end of the paper. 

\begin{figure} 
\includegraphics[width=\columnwidth]{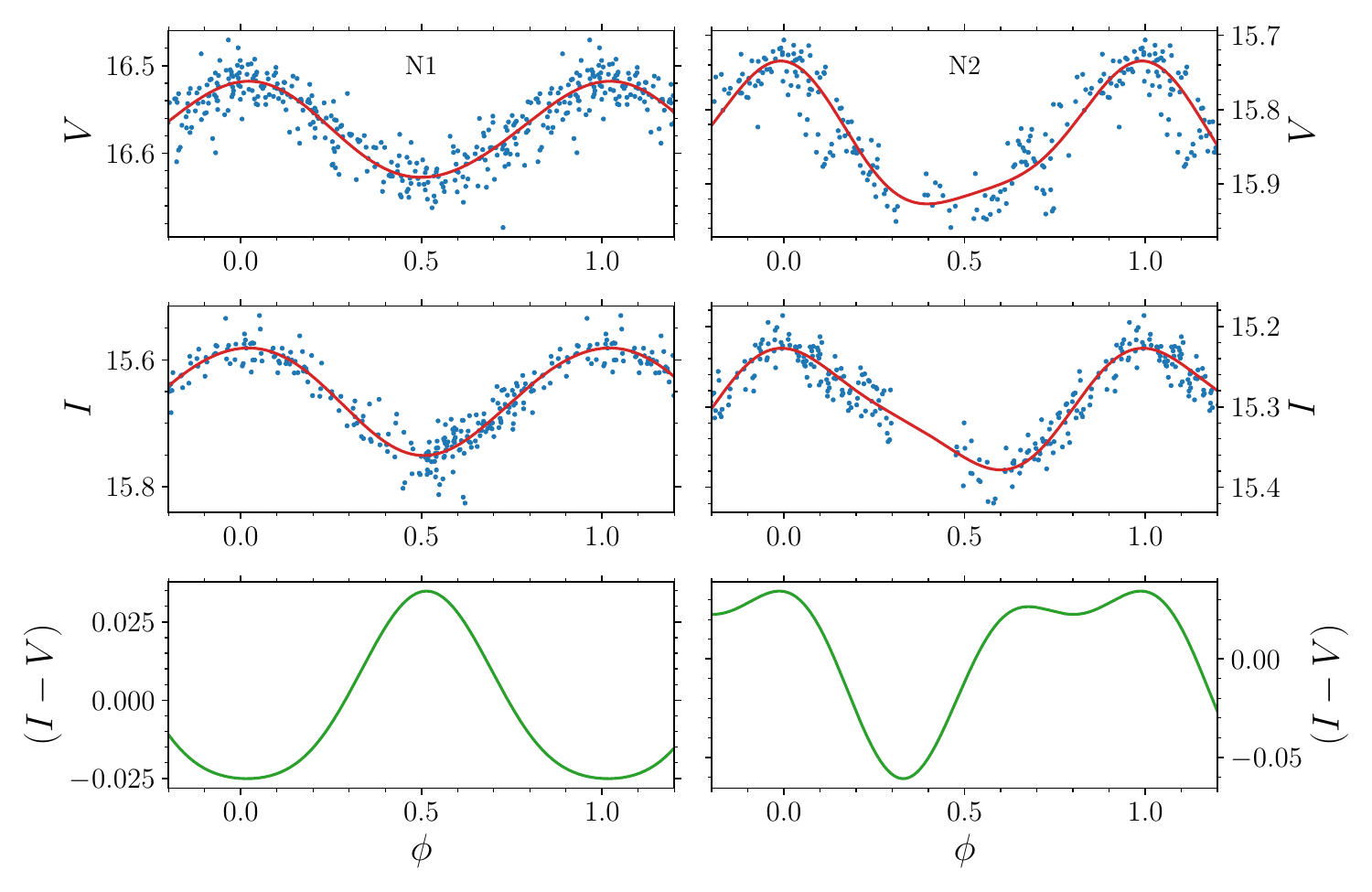}
\caption{Two newly detected variables, N1 and N2, which do not pertain to NGC 6558. Their light curves are phased with periods $0.303797$\,d and $0.342412$\,d respectively. Blue points represent observations, Fourier fits to observed data are marked in red, and green lines (bottom panel) stand for the difference between Fourier fits to illustrate colour variation.} 
\label{NEWS}
\end{figure}   

In the process we discovered two variables not reported neither in $Gaia$ nor in OGLE databases and which we identified as N1 and N2. In our membership analysis these stars were found to be field stars. Their light curves are shown in Fig. \ref{NEWS}. Judging by their periods and light curve morphology they resemble RRc variables, however we note that for N1 the amplitude in the $I-$band is a bit larger than in the $V-$band (Amp$_{V} = 0.110 \pm 0.002$\,mag and Amp$_{I} = 0.170 \pm 0.003$\,mag), which is unusual. On the other hand, N2 exhibits expected amplitude difference with amplitude in the $V$-band being larger (Amp$_{V} = 0.193 \pm 0.005$\,mag and Amp$_{I} = 0.151 \pm 0.003$\,mag).

\begin{figure*}
\includegraphics[width=17.5 cm]{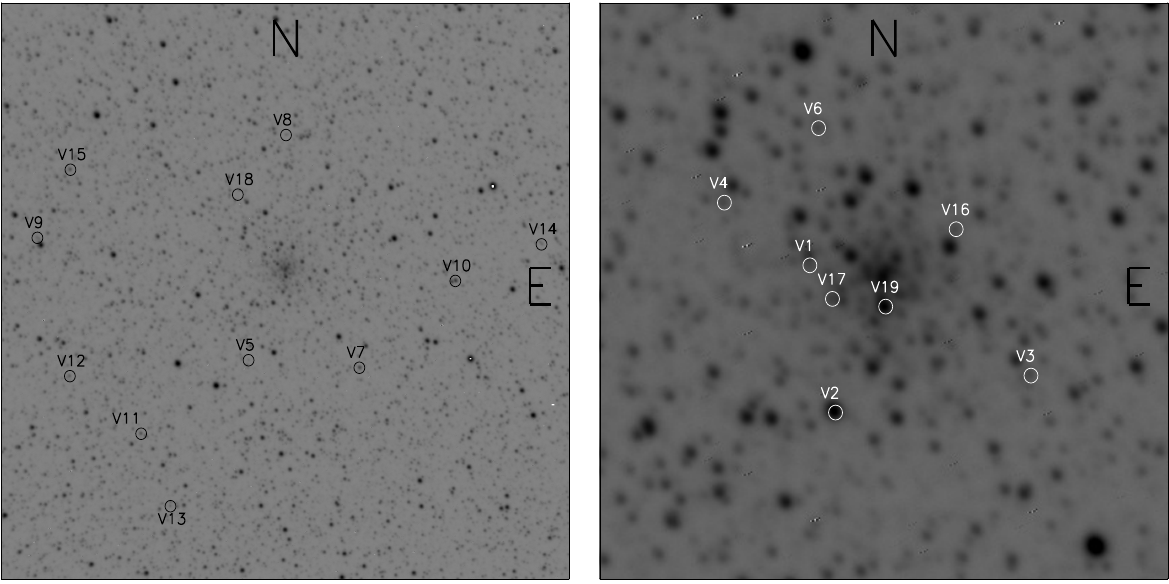}
\caption{Identification chart of the cluster member variable stars NGC 6558. The field of the left panel is about 13.2$\times$13.2 arcmin$^2$ whereas the central region in the right panel is  approximately 3.3$\times$3.3 arcmin$^2$} 
\label{CHART}
\end{figure*}

In the identification chart of Fig. \ref{CHART} we include all stars listed in the CVSGC (2016 edition) plus the two new member variables discovered in this work (V18 and V19). Note that V2 and V7 in fact do not show signs of variability, and that 10 of them are not cluster members, as indicated in Table \ref{match}.

\begin{figure}
\includegraphics[width=8.0 cm]{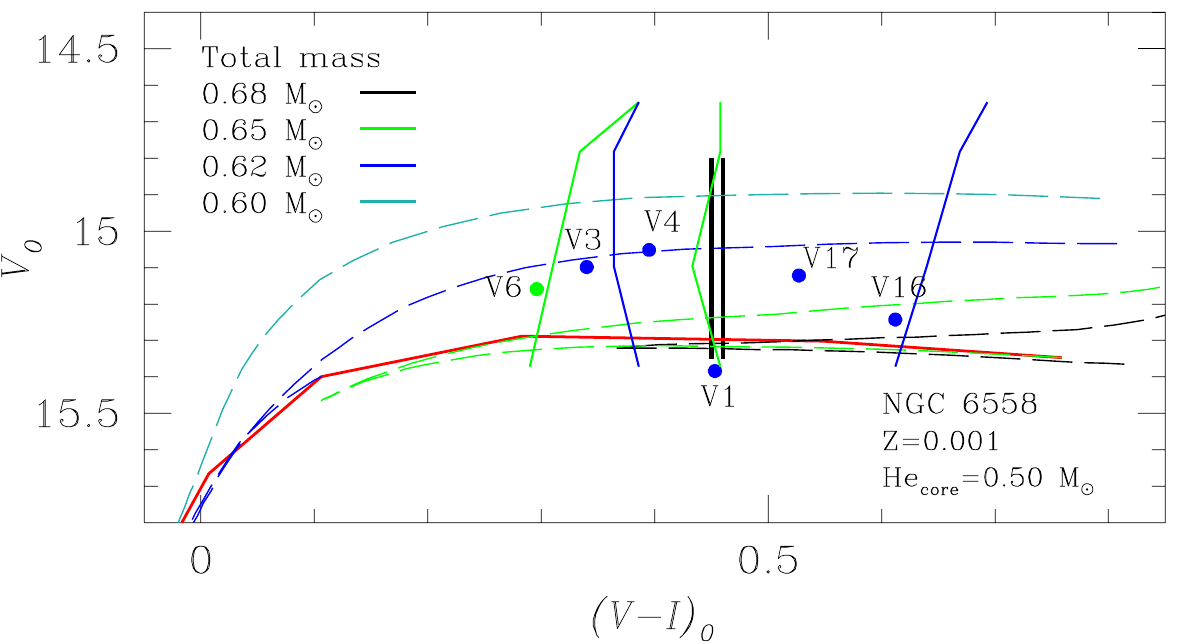}
\caption{Horizontal branch region of NGC 6558. Blue and green symbols represent RRab and RRc cluster member stars respectively. Red continuous line and vertical black lines are the ZAHB and first overtone red edge described  in the caption of Fig. \ref{CMDs}. Segmented lines are evolutionary tracks with total and core masses given in the figure legend. Blue and green vertical loci represent the instability strip borders for the fundamental and first overtone respectively \citep{Bono1994}}. See Section \ref{HBmodels} for a discussion. 
\label{HB_6558}
\end{figure} 

\begin{table*}
\footnotesize
\centering 
\caption[]{\small Fourier coefficients for RRab stars. The numbers in parentheses indicate the uncertainty on the last decimal place. The deviation parameter $D_{\mbox{\scriptsize m}}$ is given in th elast column.}
\centering                   
\begin{tabular}{lllllllllr}
\hline
Variable & $A_{0}$ & $A_{1}$ & $A_{2}$ & $A_{3}$ & $A_{4}$ & $\phi_{21}$ & $\phi_{31}$ & $\phi_{41}$ &  $D_{\mbox{\scriptsize m}}$ \\
   & ($V$ mag)  & ($V$ mag)  &  ($V$ mag) & ($V$ mag)& ($V$ mag) & & & & \\
   \hline
V1 & 16.634(2) & 0.382(2) & 0.177(2) & 0.124(2) & 0.077(2) & 3.8453(19) & 8.127(27) & 6.133(41) & 2.7\\
V3 & 16.280(3) & 0.456(4) & 0.239(4) & 0.133(4) & 0.103(4) & 4.013(23) & 8.279(38) & 6.487(48) & 4.3\\
V4 & 16.439(4) & 0.405(4) & 0.198(3) & 0.140(3) & 0.086(3) & 3.944(21) & 8.159(31) & 6.174(46) & 1.1\\
V16 & 16.579(1) & 0.128(2) & 0.052(2) & 0.007(2) & 0.006(2) & 4.497(35) & 9.557(187) & 7.813(296) & 3.7\\
V17 & 16.248(1) & 0.093(1) & 0.019(1) & 0.005(1) & 0.001(1) & 4.508(82) & 9.116(28) & 8.049(99) & 5.2\\
\hline
\end{tabular}
\label{fourier_coeffs}
\end{table*}

\begin{table*}
\footnotesize
\centering
\caption[] {\small Physical parameters for the RRab stars. The numbers in parentheses indicate the uncertainty on the last decimal place.}
\hspace{0.01cm}
 \begin{tabular}{cccccccccc}
\hline 
Variable&[Fe/H]$_{\rm ZW}$  &[Fe/H]$_{\rm UVES}$ &[Fe/H]$_{\rm DK}\dagger$ & $M_V$ & log~$T_{\rm eff}$  & log~$(L/{\rm L_{\odot}})$ & $D$ (kpc) & $M/{\rm M_{\odot}}$&$R/{\rm R_{\odot}}$\\
\hline
V1 & $-1.21(3 )$ & $-1.09(3)$ &--1.309 &0.691(3) & 3.822(8) & 1.626(1)& 9.01(1) & 0.70(7) & 4.49(1)\\
V3 & $-1.52(4)$& $-1.44(4)$&--1.994*
&0.423(6)& 3.810(9) & 1.741(2)& 8.75(3) & 0.79(8) & 5.56(2)\\
V4 & $-1.32(3)$ & $-1.20(3)$& --1.296&0.614(4) & 3.818(9) & 1.659(2)& 8.21(3) & 0.71(7) & 4.69(1) \\
V16 &--0.70 (22)*    & --0.68 (12)*   &  --1.240    & 0.573(3)& 3.807(34) & 1.663(1)& 8.98(1) & 0.52(21) & 5.63(10)\\
V17 & $-1.37(5)$ & $-1.26(5)$&--1.257 &0.535(1) & 3.786(13) & 1.692(1)& 7.84(3) & 0.63(10) & 5.94(2) \\
\hline
Weighted Mean & $-1.33(2)$ & $-1.20(2)$ & --1.280&0.569(8) & 3.813(1) & 1.670(1) & 8.47(4) & 0.69(3) & 5.66(2)\\
$\sigma$&  $\pm$0.11  &  $\pm$0.13  & $\pm$0.028&$\pm$0.089&$\pm$0.012&$\pm$0.039 & $\pm$0.46 &$\pm$0.09&$\pm$0.57\\
\hline 
\end{tabular}

\quad 
$\dagger$ Iron values from \citet{Dekany2021}. The extreme value of V3 is likely due to blending and was not averaged.  *Not included in the average of [Fe/H] due to uncertain $\phi_{31}$.
\label{fisicos}
\end{table*}


\section{The Colour-Magnitude diagram}
\label{sec:CMD}
In the CMD of Fig.~\ref{CMDs} we included all the variables in the field of NGC 6558 coded as described in the caption. All stars whose variability has been confirmed are plotted using their intensity-weighted
means $<V>_0$ and corresponding colour $<V>_0 - <I>_0$. Filled and empty symbols were used for members and either field stars or of unknown status due to the lack of proper motions in the $Gaia$-DR3 database. 

All member variables occupy the expected regions according to their variable type. We note the distribution of RRab stars relative to the empirical first overtone red edge (FORE) indicated in the DCM by the vertical black lines in the HB. See Fig. \ref{HB_6558} for an expanded version of the HB region. There is evidence that at least two RRab stars V3 and V4, are sitting to the left of the FORE, i.e. in the bimodal or "either-or" region, shared by fundamental and first overtone mode pulsators. Both stars show signs of amplitude modulations. The period analysis of V4 (RRLYR-14866) using \texttt{Period04} \citep{Lenz2004Period04} shows a periodic modulation of the light curve equal to $P_{\rm BL} = 26.73 \pm 0.01$\,day \citep[known as the Blazhko effect,][]{Blazhko1907}. V1, fundamental mode variable, seems a border case between blue and red parts of the instability strip.

\begin{figure}
\includegraphics[width=6.0 cm]{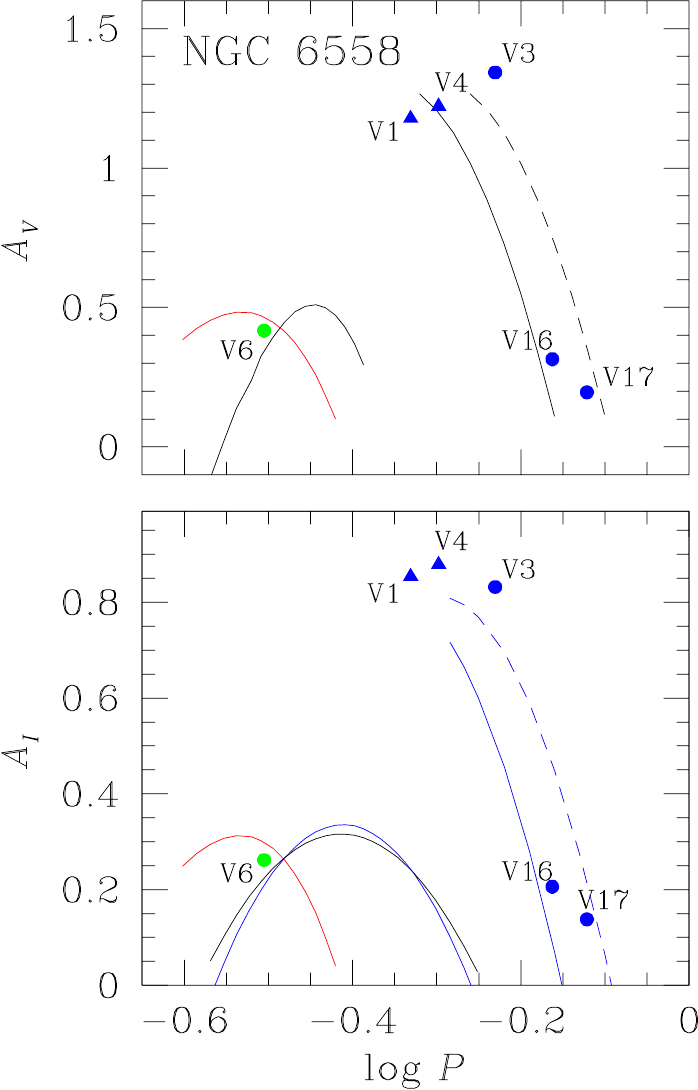}
\caption{Period-Amplitude diagram. Continuous and segmented loci to the right are indications of Oo I and Oo II types respectively for the RRab stars distribution \citep{Cacciari2005} and \citep{Kunder2013a}: likewise are the red and black parabolas for RRc stars \citep{Kunder2013b} and \citep{Yepez2020}. The distribution of stars do not favour neither Oo I nor Oo II types. The RRab variables V1 and V4 were found to be cluster members in our analysis, but were given a low membership probability by the analysis of \citet{Vasiliev2021}}. See discussion in section \ref{sec:CMD}. 
\label{Bailey}
\end{figure}

According to \citet{Arellano2019}(their figure 8) or \citet{Deras2022} (their figure 7), the presence of fundamental mode RRab stars in the bimodal region is an exclusive characteristic of some Oo I clusters, and does not occur in Oo II type clusters. This comment is of relevance considering that the Oosterhoff type of NGC 6558 cannot be assessed clearly: in the Bailey's diagram of Fig.  \ref{Bailey} member stars are few and their distribution do not decant in favour of either Oo I or OoII sequences; the average period of the 5 member RRab stars is 0.60 $\pm$ 0.04 day which is the border between Oostherhoff types. The empirical border [Fe/H]$_{\rm ZW} \sim -1.5$ between the metal poor Oo II and the metal richer Oo I \citep{Arellano2024}, and the value of [Fe/H]$_{\rm ZW} =-1.33 \pm 0.11$ obtained from the Fourier decomposition of the best observed members (V1, V3, V4 and V17). The above average period and metallicity, when plotted on the plane [Fe/H] vs <P$_{ab}$> of figure 5 of \citet{Catelan2009}, places the cluster in the Oosterhoff gap, suggesting that NGC 6558 is of the intermediate Oosterhoff type or Oo-Int. We note that in the work by \citet{Catelan2009}, NGC 6558 was classified as Oo I due to the lower average pulsation period for associated fundamental mode pulsators.

A revision of the membership probabilities assigned by \citet{Vasiliev2021} for stars in the field of NGC 6558, confirm the membership status of V3, V16 and V17, but assigns very low probabilities to V1 and V4. If these two stars are not considered members the average <P$_{ab}$> becomes 0.677 d, which would only highlight NGC 6558 as a peculiar cluster in the [Fe/H] vs <P$_{ab}$> plane, i.e. too high a metallicity for its average period. Two clusters with a similar property are the moderate metal-rich Oo III clusters NGC 6388 and NGC 6441 which, in spite their extended HB blue tails \citep{Piotto1997,Rich1997}, they display a prominent red-clump which is absent in NGC 6558.

In our opinion and based on our membership analysis, the five RRab members lead to a value of <P$_{ab}$>=0.60$\pm$0.04, which puts the cluster in the Oosterhoff gap in the [Fe/H] vs <P$_{ab}$> plane and NGC 6558 should be considered of an Oo-int nature.

\section{Fourier light curve decomposition of member RR Lyrae}
\label{Fourier}

We identified five RRab member stars of the cluster. In this section we perform a Fourier decomposition of their light curves and use their Fourier parameters and ad-hoc semi-empirical calibrations to estimate their metallicity and distance. For briefness we will not repeat here the description of this approach since it has been thoroughly described by \citet{Arellano2010} and summarized by \citet{Arellano2024}.

In Table \ref{fourier_coeffs} the Fourier coefficients of the five member RRab are listed as it is their consistency parameter $D_m$ defined by \citep{Jurcsik1996}. According to these authors, the [Fe/H] calibration for RRab stars is applicable for values of $D_m < 3.0$. We have relaxed this criterion a bit to 
$D_m < 5.0$ to increase the possibilities of our sample.

We recall that the iron abundance value obtained photometrically from this calibration is given into the scale of Jurcskic and Kov\'acs, [Fe/H]$_{\rm JK}$, which can be converted into the scale of \citet{Zinn1984} via [Fe/H]$_{\rm JK}$ = 1.431[Fe/H]$_{\rm ZW}$ + 0.88 \citep{Jurcsik1996} and in turn it can be transformed into the spectroscopic scale of \citet{Carretta2009} via [Fe/H]$_{\rm UVES}$= $-0.413$ + 0.130~[Fe/H]$_{\rm ZW} - 0.356$~[Fe/H]$_{\rm ZW}^2$. In Table \ref{fisicos} we report the mean physical parameter obtained for the 5 RRab stars. For comparison, we have included in column 4 of Tab. \ref{fisicos} the iron values estimated by \citet{Dekany2021} [Fe/H]$_{\rm DK}$, for the same RR Lyrae in NGC 6558. These authors used the photometric Fourier decomposition approach considering newly calculated calibrations of the $I$-band light curve parameters. The average [Fe/H]$_{\rm DK}$  matches well our results for [Fe/H]$_{\rm UVES}$. 

\section{On the distance to NGC 6558}
\label{Sec:distance}

The P-L ($I$) calibration for RR Lyrae stars can be used to produce independent estimates of the distance. We have employed two calibrations available in the literature. \citet{Catelan2004} calculated the equation $M_I = 0.471-1.132~ {\rm log}~P +0.205~ {\rm log}~Z$, with ${\rm log}~Z = [M/H]-1.765$; $[M/H] = \rm{[Fe/H]} - \rm {log} (0.638~f + 0.362)$ and log~f = [$\alpha$/Fe], from where we adopted [$\alpha$/Fe]=+0.4 \citep{Sal93}. For the calculation, we have adopted the value of the metallicity [Fe/H]$_{\rm UV}$ = -1.20, obtained from the Fourier approach (see Table \ref{fisicos}, and the average $E(B-V)$= 0.40. We applied the above calibration to the five RRab stars in Table \ref{fisicos} and the RRc star V6. The period of the latter was fundamentalized using the ratio $P_1/P_0$ = 0.75. The resulting mean distance was $8.10 \pm 0.22$ kpc, which is in good agreement with the result from the Fourier approach of $8.47 \pm 0.46$ kpc.

On the other hand, a recent P-L ($I$) calibration calculated by \citet{Prudil2024} using data published by the \textit{Gaia}-DR3 is of the form $M_I = 0.197-1.292~ {\rm log}~P +0.196~{\rm[Fe/H]}$. When this equation is applied to the six cluster members RRab and RRc, a mean distance of 7.72 $\pm$ 0.20 kpc was found. We note that for the treatment of reddening and differential reddening, we used reddening maps and laws from \citet{Schlegel1998} and \citet[][see Section~\ref{differential}]{AlonsoGarcia2012}.

With the P-L relations in the aforementioned paper, \citet{Prudil2024} derived reddening maps and reddening laws for the following four colours: $(J-K_{\rm s}), (I-K_{\rm s}), (V-I)$ and $(G_{\rm BP}-I)$. Using the newly reddening maps and reddening laws, solely based on individual RR~Lyrae variables toward the Galactic bulge, we have obtained their distances,
using a similar approach as used by \citet{Prudil2019}. The results are shown in Table \ref{distance_Zdenek}. In the last column, we have averaged the results for colours  $(J-K_{\rm s})$and $(V-I)$ which in turn lead to a grand average of 8.73$\pm$0.29. Comparing these results with those in Table \ref{tab:distance} we point to the agreement, within the respective uncertainties, with the distance obtained from the Fourier light curve decomposition approach.

It should be noted that in their compilation of globular cluster distances, Bumgardt et al. 2023\footnote{\texttt{https://people.smp.uq.edu.au/HolgerBaumgardt/globular/}} the distances determined for NGC 6558 prior to 1998, tend to be smaller than 8 kpc and the authors report an average of $7.79 \pm 0.18$ kpc. In their paper \citet{Baumgardt2021} the reported mean distance is $7.47 \pm 0.29$ kpc. In either case our calculations via the Fourier approach and the P-L($I$) relationship render a distance a bit larger than 8 kpc, as stated above. For a clearer reference, in Table \ref{tab:distance} we summarize all the above distances and methods

\begin{table}
\footnotesize
\caption{Summary of distance estimates of NGC 6558. }
\centering
\begin{tabular}{ll}
\hline
$D$  &  source \\
 kpc     &  \\
\hline
 8.47$\pm$0.46   & This work RR Lyrae Fourier decomposition \\
 8.10$\pm$0.22   & P-l ($I$) \citep{Catelan2004} \\
 7.72$\pm$0.20   & P-l ($I$) \citep{Prudil2024} \\
7.79$\pm$0.18& Baumgardt et al. (2023) compilation\\ 
7.47$\pm$0.29& \citep{Baumgardt2021}\\
\hline
\end{tabular}
\label{tab:distance}
\end{table}

\begin{table*}
\footnotesize
\caption{Cluster member RR Lyrae distances (derived in Prudil et al. 2024, submitted) based on P-L relations of \citet{Prudil2024}.}
\centering
\begin{tabular}{ccccccc}
\hline
	\hline
ID Star& ID OGLE	& $D_{J-K_{\rm s}}$ (kpc)  & $D_{I-K_{\rm s}}$ (kpc) & $D_{V-I}$ (kpc)   & $D_{G_{\rm BP}-I}$ (kpc) & Avg [$D_{J-K_{\rm s}}$;$D_{V-I}$] (kpc) \\
	\hline
 V1 & RRLYR-14886 & 7.73 & 7.68 & 8.77 & 9.40& 8.25 \\
 V3 & RRLYR-14929 & 9.00 & 8.93 & 8.87 & 7.22& 8.94 \\
 V4 & RRLYR-14866 & 9.24 & 9.28 & 9.00 & 8.57& 9.12 \\
 V6 & RRLYR-14888 & 8.77 & 8.89 & 8.61 & 8.14& 8.83 \\
 V16& RRLYR-14912 & 8.86 & 8.85 & 8.10 & 7.93& 8.48 \\
 V17& RRLYR-14892 & 9.01 & 9.02 & 8.53 & 8.49& 8.77 \\
  \hline
 Average &&&&&&	8.73$\pm$0.29 \\
 \hline
\end{tabular}
\label{distance_Zdenek}
\end{table*}

\section{Modelling the Horizontal Branch}
\label{HBmodels}

There is a clear correlation between the HB extension to the blue and the metallicity among globular clusters \citep{Sandage1960}; the lower the metallicity, the bluer the HB. The temperature of a HB star depends both on the metallicity (lower metallicity produces lesser opacities and then more compact, hotter, and bluer shells), but also on the mass of a H-rich shell. It was noted by \citet{KPS2005} that by comparing models of a given metallicity but different shell masses it can be shown that lower shell masses also produce bluer HB stars. 

The role of mass loss during the He-flashes events at the RGB upon the colour of a He-core burning star settling on the HB, has been illustrated by \citet{SilvaAguirre2008}. The more mass is lost in the RGB the bluer the star will be on the ZAHB when the helium core gets ignited. 

To model the RR Lyrae region of NGC 6558 we used the code employed by \citet{KPS2005} that uses a modified Reimers law with $\eta$ = $0.8 \times 10^{-13}$. 
This original evolution code 
developed by 
\citet[][]{Eggleton1971, Eggleton1972, Eggleton1973}
 and further improved and thoroughly 
tested by \citet[][]{Pols1997, Pols1998}  and \citet{KPS1997}. We here use abundances for Z=0.001, equivalent to [Fe/H]~$=-1.3$\,dex, similar to the metallicity found
for NGC 6558.

In Fig. \ref{HB_6558} the five cluster member RR Lyrae are shown on an expanded version of the HB region. The stars are well represented by models with a core mass of 0.50 $M_{\odot}$ and a range of shell mass between 0.10 and 0.18 $M_{\odot}$. The main-sequence progenitor of these stars is a 0.82-0.86 $M_{\odot}$ star that reached the RGB after approximately 13 and 11 Gyrs respectively, where it lost some 20 to 30\% of its mass before evolving to the ZAHB.

In older globular clusters, HB stars have developed from slightly lower mass stars on the main sequence. Since the degenerate Helium-core needs in all cases $\sim$ 0.5 $M_{\odot}$ to start the central He-burning, then the resulting H-rich shell mass are smaller and bluer. We have argued \citep{Arellano2020} that the range of colour on the HB population, larger than expected from a suitable model and their moves on the HRD \citep{VandenBerg1990}, can be explained by slight variations of the mass loss on the RGB, which produces certain shell-mass range of HB stars in the same globular cluster. A moderate star-to-star variation of mass-loss would make a simple explanation for the extended colour distributions on the HB. We have also speculated that mass-loss in the  RGB may be modulated by the presence of magnetic fields in red giants \citep{Konstantinova2013}.

\section{Summary and Conclusions}
\label{sec:Summ}
Since the stellar field in the Galactic bulge is densely populated, the number of variable stars in the field of a Galactic bulge cluster can seem very large, however, it is very likely that many of those variables do not pertain to the cluster. That such is the case of NGC 6558 has been demonstrated in this work. In the  Catalogue of Variable Stars in Galactic Globular Clusters, OGLE and $Gaia$-DR3 data, we identified 78 stars in the field  of NGC 6558. However a proper motion  analysis and a thorough exploration of this variables on the CMD of the cluster  shows that only 9 of them seem to be truly cluster members.

We have then Fourier decomposed the light curves of the five RRab members to estimate the mean distance and metallicity of the cluster as
$8.47 \pm 0.46$ kpc and 
[Fe/H]$_{\rm ZW} =-1.33 \pm 0.11$,  or  in the spectroscopic scale of \citet{Carretta2009} [Fe/H]$_{\rm UVES} =-1.20 \pm 0.13$.

A cross identification of all variables in the field of the cluster between their OGLE and $Gaia$-DR3 numbers is provided and \emph{VI} light curves are displayed and made available in the Centre de Donnés astronomiques de Strasbourg database (CDS). Their ephemerides, AR and DEC, and mean \emph{VI} magnitudes and amplitudes are also tabulated.

\noindent
{\bf Acknowledgements.} AAF is grateful to the Indian Institute of Astrophysics, and to the European Southern Observatroy for warm hospitality during the preparation of this work. AAF also thankfully acknowledges the sabbatical support granted by the
program PASPA of the DGAPA-UNAM.  We have been benefited from the support of DGAPA-UNAM through projects IG100620 and IN103024.
We have made an extensive use of the SIMBAD and ADS services, for which we are
thankful.

{\bf Data Availability:} The data underlying this article shall be available in electronic form in the Centre de Donnés astronomiques de Strasbourg database (CDS), and can also be shared on request to the corresponding author.

\bibliographystyle{mnras}
\bibliography{N6558_MNRAS} 

\appendix
\label{Appendix}

\section{Variables in the field of NGC6558}

In Table \ref{match} the variables found in the field of the cluster are listed with the cross identification from their sources, i.e. the CVSGC \citep{Clement2001}, OGLE \citep{Soszynski2013}, \citep{Soszynski2014} and $Gaia$-DR3 \citep{Gaia2018}. The membership status is indicated by an asterisk in column 3. In Table \ref{FIELD_VAR} there are reported the intensity weighted mean magnitudes, amplitudes and epochs for all variables in the field, regardless they are members or not.

\begin{table*}
\caption{Crossmatch between OGLE and $Gaia$-DR3 identifications of variable stars in the field of NGC 6558.}
\begin{center}
\label{match}
	\begin{tabular}{ccl|ccl}
			\hline
			ID Star& ID OGLE	& ID GAIA     & ID Star & ID OGLE   & ID GAIA \\
            &(OGLE-BLG+)&     &  & (OGLE-BLG+)	& \\
	\hline
 V1 & RRLYR-14886 & 4048864059416545280* & O22 & ECL-342853 & 4048864059416638208 \\
V2 & ---------------  & 4048864432989883264* & O23 & ECL-342957 & 4048864506093328768 \\
V3 & RRLYR-14929 & 4048864265574910592* & O24 & ECL-343193 & 4048863750138723840 \\
V4 & RRLYR-14866 & 4048864025056821888* & O25 & ECL-343374 & 4048860692161846400\\
V5 & RRLYR-14867 & 4048865257631129856* & O26 & ECL-343590 & 4048863784560136832 \\
V6 & RRLYR-14888 & 4048863814514496384* & O27 & ECL-343705 & 4048860790969550208 \\
V7 &  ---------------  & 4048864403013942528 & O28 & ECL-343910 & 4048860825368698624 \\
V8 & RRLYR-14897 & 4048860692161845760 & O29 & ECL-343918 & 4048864467349644928 \\
V9 & RRLYR-14732 & 4048841480684585600 & O30 & ECL-344196 & 4048864368615454336 \\
V10 & --------------- & 4048861306252440832 & O31 & ECL-344354 & 4048864192625493120 \\
V11 & --------------- & 4048865223264032896 & O32 & ECL-344375 & 4048864162495551616 \\
V12 & --------------- & 4049052758715416320 & O33 & ECL-344488 & 4048861207517918080 \\
V13 & RRLYR-14809 & 4049053308471439744 & O34 & ECL-344503 & 4048864604944711296 \\
V14 & RRLYR-15057 & 4048862513227644032  & O35 & ECL-344522 & 4048860550488673408 \\
V15 & --------------- & 4048841102727354368 & O36 & ECL-344792 & 4048860550488742016 \\
V16 & RRLYR-14912 & 4048864226982967168* & O37 & ECL-344831 & 4048864295704788736 \\
V17 & RRLYR-14892 & 4048864055040290176* & O38 & ECL-345018 & 4048861344956949504 \\
V18 (O20) & ECL-342710 & 4048863711547338752* & O39 & ECL-345114 & 4048861173199302528 \\
V19 (G5) & --------------- & 4048864226831441920* & O40 & ECL-345412 & 4048861173157993600 \\
O1 & DSCT-08146 &   4048864089504042880 & O41 & ECL-345522 & 4048861168966369152 \\
O2 & DSCT-08222 & 4048863848986229504 & O42 & ECL-345774 & 4048860962807893632 \\
O3 & DSCT-14210 & 4048864746611547008 & O43 & ECL-346064 & 4048861306407735424 \\
O4 & ECL-340840 & 4048864780971419776 & O44 & ELL-021823 & 4048860829601924480 \\
O5 & ECL-341024 & 4048864707979313408 & O45 & ELL-021837 & 4048864196855360640 \\
O6 & ECL-341028 & 4048840488628088320 & O46 & ELL-021862 & 4048861168854110720 \\
O7 & ECL-341420 & 4048864918370561280 & O47 & RRLYR-14873 & 4048864059416651392 \\
O8 & ECL-341506 & 4048840419908456320 & O48 & DSCT-08146 & ---------------------- \\
O9 & ECL-341526 & 4048863956337442432 & O49 & ECL-346111 & 44048866254056274176 \\
O10 & ECL-341544 & 4048864849690854016 & O50 & RRLYR-14709 & ---------------------- \\
O11 & ECL-341790 & 4048863990697244800 & O51 & DSCT-08190 & 4048840007591268224 \\
O12& ECL-341801 & 4048864845419442432 & O52 & ECL-339704 & 4048841480795839872 \\
O13 & ECL-341917 & 4048865326382558848 & O53 & ECL-344368 & 4048866155360805632 \\
O14 & ECL-342132 & 4048864089504046976 & O54  & ECL-345391 & 4048866116773356416 \\
O15 & ECL-342151 & 4048864089504247680 & O55  & ECL-346668 & 4048865983561808128 \\
O16 & ECL-342461 & 4048864025017002368 & O56  & -DSCT-14242 & ---------------------- \\
O17 & ECL-342462 & 4048864884050514432 & N1  & --------------- & ----------------------- \\
O18 & ECL-342540 & 4048864025056864896 & N2  & ----------------- & 4048866155360747776\\
O19 & ECL-342585 & 4048863608466744320 & G8 & --------------- & 4048861173198138752 \\
 O21 & ECL-342716 & 4048864025057644416 & G10 & ----------------- & 4048863608355972480\\
\hline
\end{tabular}
\end{center}

\quad $*$
Cluster member star
\end{table*}

\begin{table*}
\caption{General data for variables in the field of the cluster. The large majority are not cluster member stars. All periods and epochs were estimated in the present work.}
\begin{center}
\label{FIELD_VAR}
	\begin{tabular}{ccccccccccc}
			\hline
			ID & Variable & $<V>$      & $<I>$  & $A_V$  & $A_I$  & P & Epoch & RA & DEC & m/f/u \\

			 & Type	&(mag)     & (mag) &  (mag)  &  (mag) & (days) & (+245 0000.) & (2000.) & (2000.) &  \\
           
	\hline
V1 & RRab & 16.634 & 15.662 & 1.154 & 0.854 & 0.466592 & 2869.64820 & 18:10:15.59 &$-$31:45:54.10 & m \\
V2 & CST & 13.553 & 11.307 & ------ & ------ & ------ & ------ & 18:10:16.26 & $-$31:45:03.46 & m \\
V3 & RRab & 16.280 & 15.376 & 1.285 & 0.859 & 0.588021 & 2574.51730 & 18:10:21.64 & $-$31:45:15.57 & m \\
V4 & RRab & 16.439 & 15.543 & 1.262 & 0.899 & 0.503936 & 8334.65550 & 18:10:13.27 & $-$31:46:16.03 & m \\ 
V5 & RRab & 16.836 & 15.610 & 0.863 & 0.520 & 0.748679 & 5749.55520 & 18:10:13.48 & $-$31:43:45.12 & f \\
V6 & RRc & 16.466 & 15.629 & 0.572 & 0.323 & 0.312489 & 2563.53700 & 18:10:15.88 & $-$31:46:41.44 & m  \\
V7 & CST & 14.603 & 13.632 & ------ & ------ & ------ & ------ & 18:10:25.61 & $-$31:43:33.19 & f  \\
V8 & RRab & 16.478 & 15.505 & 1.308 & 0.790 & 0.574526 & 2916.61750 & 18:10:17.87 & $-$31:48:55.31 & f \\
V9 & RRab & 16.514 & 15.634 & 1.322 & 0.867 & 0.454453 & 5453.62460 & 18:09:50.47 & $-$31:46:37.28 & f \\
V10 & L & 13.978 & 11.831 & ------ & ------ & ------ & ------ & 18:10:36.26 & $-$31:45:31.36 & f \\
V11 & L & 15.230 & 12.791 & ------ & ------ & ------ & ------ & 18:10:01.62 & $-$31:42:05.43 & f \\
V12 & L & 16.037 & 10.689 & ------ & ------ & ------ & ------ & 18:09:53.86 & $-$31:43:25.82 & f \\
V13 & RRc & 16.431 & 15.554 & 0.384 & 0.283 & 0.360992 & 2563.53700 & 18:10:04.75 & $-$31:40:24.89 & f \\
V14 & RRc & 16.412 & 15.496 & 0.514 & 0.386 & 0.265041 & 2808.87910 & 18:10:45.76 & $-$31:46:20.29 & f \\
V15 & L & 15.230 & 11.944 & ------ & ------ & ------ & ------ & 18:09:54.16 & $-$31:48:10.89 & f \\
V16 & RRab & 16.579 & 15.413 & 0.386 & 0.256 & 0.687550 & 2135.69400 & 18:10:19.63 & $-$31:46:05.97 & m \\
V17 & RRab & 16.250 & 15.255 & 0.229 & 0.149 & 0.755803 & 2432.77310 & 18:10:16.21 & $-$31:45:42.38 & m \\
V18 (O20) & ECL & 19.549 & 18.130 & 0.543 & 0.569 & 0.406042 & 7000.39400 & 18:10:12.50 & $-$31:47:33.78 & m \\
V19 (G5) & SR/L & 13.7550 & 11.527 & ------ & ------ & 346.185375 & 6911.53619 & 18:10:17.67 & $-$31:45:39.42 & m \\
N1  & RRc & 16.573 & 15.663 & 0.189 & 0.188 & 0.303797 & 8376.55965 & 18:10:40.99 & $-$31:39:48.18 & ------ \\
N2  & RRc & 16.067 & 15.311 & 0.257 & 0.219 & 0.342412 & 8692.56056 & 18:10:25.66 & $-$31:41:59.10 & f \\
G8 & ------ & 17.198 & 13.655 & 1.674 & 0.977 & 70.985000 & 8692.54054 & 18:10:30.45 & $-$31:46:34.60 & f \\
G10 & ------ & 15.727 & 14.752 & 0.108 & 0.043 & 0.178146 & 8698.51780 & 18:10:11.80 & $-$31:48:22.30 & f \\
O1 & DSCT & 17.679 & 16.887 & 0.426 & 0.227 & 0.072299 & 7000.06928 & 18:10:06.54 & $-$31:45:38.03 & f \\
O2 & DSCT & 17.905 & 16.589 & 0.323 & 0.194 & 0.105038 & 7000.01133 & 18:10:17.53 & $-$31:45:50.60 & u \\
O3 & DSCT & 17.494 & 16.610 & ------ & ------ & 0.068512 & 7000.02705 & 18:10:03.57 & $-$31:46:07.34 & f \\
O4 & ECL & 18.056 & 17.013 & 0.547 & 0.380 & 1.734955 & 5351.90512 & 18:09:59.66 & $-$31:45:25.98 & f \\
O5 & ECL & 16.617 & 15.776 & 0.504 & 0.454 & 1.018260 & 7000.91980 & 18:10:01.10 & $-$31:45:44.82 & f  \\
O6 & ECL & 15.050 & 14.183 & 0.173 & 0.166 & 0.442419 & 7000.22190 & 18:10:01.12 & $-$31:47:15.03 & f  \\
O7 & ECL & 18.284 & 17.403 & 0.362 & 0.411 & 0.442263 & 7000.08360 & 18:10:03.85 & $-$31:44:42.05 & f \\
O8 & ECL & 17.792 & 17.017 & 0.233 & 0.269 & 0.426111 & 7000.00280 & 18:10:04.34 & $-$31:48:04.62 & f \\
O9 & ECL & 19.690 & 18.397 & 0.302 & 0.284 & 0.393296 & 7000.27860 & 18:10:04.45 & $-$31:47:15.00 & u \\
O10 & ECL & 18.543 & 17.386 & 0.310 & 0.324 & 0.442397 & 7000.29750 & 18:10:04.63 & $-$31:44:50.95 & f \\
O11 & ECL & 20.183 & 18.857 & 0.493 & 0.443 & 0.380086 & 7000.32450 & 18:10:06.24 & $-$31:46:10.00 & u \\
O12 & ECL & 19.728 & 18.400 & 0.372 & 0.352 & 0.316283 & 7000.12300 & 18:10:06.34 & $-$31:44:48.65 & f \\
O13 & ECL & ------ & 19.290 & ------ & 0.665 & 0.349780 & 7000.02210 & 18:10:07.02 & $-$31:42:53.30 & u \\
O14 & ECL & 20.696 & 19.380 & 0.521 & 0.471 & 0.340083 & 7000.25920 & 18:10:08.35 & $-$31:45:49.62 & f \\
O15 & ECL & 20.059 & 18.807 & 0.677 & 0.691 & 0.344149 & 7000.14650 & 18:10:08.46 & $-$31:45:44.71 & u \\
O16 & ECL & 20.377 & 18.956 & 0.486 & 0.464 & 0.332756 & 7000.29330 & 18:10:10.76 & $-$31:46:10.60 & f \\
O17 & ECL & 20.692 & 19.147 & 0.439 & 0.599 & 0.331755 & 7000.20540 & 18:10:10.53 & $-$31:44:25.00 & u \\
O18 & ECL & 18.459 & 17.418 & 0.273 & 0.321 & 0.458385 & 7000.13760 & 18:10:11.10 & $-$31:46:19.52 & f  \\
O19 & ECL & 19.884 & 17.033 & ------ & 0.696 & 4.065424 & 7001.22200 & 18:10:11.47 & $-$31:48:48.83 & f \\  
O21 & ECL & 18.876 & 17.581 & ------ & 0.349 & 0.751831 & 7000.27140 & 18:10:12.54 & $-$31:45:53.50 & f \\
O22 & ECL & 18.999 & 17.379 & 0.358 & 0.435 & 3.808711 & 7000.09620 & 18:10:13.26 & $-$31:45:22.93 & f \\ 
O23 & ECL & 18.535 & 17.657 & 0.166 & 0.210 & 0.359152 & 7000.20550 & 18:10:14.00 & $-$31:44:19.35 & f  \\
O24 & ECL & 18.753 & 17.451 & 0.217 & 0.193 & 0.509554 & 7000.04880 & 18:10:15.82 & $-$31:47:50.00 & f \\
O25 & ECL & 19.250 & 18.133 & 0.306 & 0.289 & 0.365249 & 7000.23660 & 18:10:17.06 & $-$31:49:05.86 & f \\
O26 & ECL & ------ & 18.955 & ------ & 0.574 & 0.333312 & 7000.20130 & 18:10:18.46 & $-$31:46:45.40 & u \\
O27 & ECL & 18.177 & 17.238 & 0.146 & 0.521 & 2.618627 & 7000.27960 & 18:10:19.40 & $-$31:47:40.97 & f \\
O28 & ECL & 19.659 & 18.546 & 0.206 & 0.335 & 0.295708 & 7000.01300 & 18:10:20.84 & $-$31:47:18.12 & f \\  
O29 & ECL & 15.784 & 14.951 & 0.017 & 0.039 & 0.503074 & 7000.50200 & 18:10:20.93 & $-$31:44:35.63 & f \\
O30 & ECL & 20.220 & 18.930 & 0.749 & 0.700 & 0.373394 & 7000.02310 & 18:10:22.80 & $-$31:44:23.90 & u \\
O31 & ECL & 19.239 & 17.887 & 0.244 & 0.313 & 1.363473 & 7001.22150 & 18:10:23.91 & $-$31:45:43.19 & f \\
O32 & ECL & 14.242 & 13.470 & 0.404 & 0.402 & 0.535965 & 7000.22590 & 18:10:24.09 & $-$31:46:08.91 & f \\
O33 & ECL & 18.992 & 18.075 & 0.291 & 0.317 & 0.580938 & 7000.44860 & 18:10:24.95 & $-$31:47:08.20 & f \\
O34 & ECL & 20.253 & 18.315 & ------ & 0.524 & 1.720336 & 7001.19860 & 18:10:25.11 & $-$31:43:12.64 & f \\
O35 & ECL & 19.605 & 18.351 & 0.654 & 0.639 & 6.403002 & 7003.56750 & 18:10:25.25 & $-$31:48:51.00 & f \\
O36 & ECL & 20.372 & 18.979 & 0.770 & 0.521 & 0.290533 & 7000.19830 & 18:10:27.16 & $-$31:49:02.40 & u \\  
O37 & ECL & 17.569 & 16.589 & 0.786 & 0.911 & 1.613186 & 7000.58810 & 18:10:27.48 & $-$31:44:53.58 & f \\
O38 & ECL & 19.260 & 17.887 & 0.142 & 0.187 & 0.319237 & 7000.31290 & 18:10:28.62 & $-$31:45:40.46 & f \\
O39 & ECL & 20.669 & 19.249 & 0.806 & 0.694 & 0.246369 & 7000.22250 & 18:10:29.30 & $-$31:46:55.10 & u \\
O40 & ECL & 17.635 & 16.641 & 0.307 & 0.331 & 0.519652 & 7000.00290 & 18:10:31.42 & $-$31:47:02.82 & f \\

\hline
\end{tabular}
\end{center}

\quad $*$
m- cluster member, f- field star, u- unknown due to lack of proper motion
\end{table*}

\begin{table*}
\addtocounter{table}{-1}
\caption{Continue}
\begin{center}
\label{match2}
	\begin{tabular}{ccccccccccc}

 			\hline
			ID & Variable & $<V>$      & $<I>$  & $A_V$  & $A_I$  & P & Epoch & RA & DEC & m/f/u \\

			 & Type	&(mag)     & (mag) &  (mag)  &  (mag) & (days) & (+245 0000.) & (2000.) & (2000.) &  \\
           
	\hline
O41 & ECL & 20.372 & 18.486 & 0.620 & 0.541 & 0.454164 & 7000.10650 & 18:10:32.03 & $-$31:47:07.00 & f \\
O42 & ECL & 18.879 & 17.705 & 0.243 & 0.292 & 0.429635 & 7000.23460 & 18:10:33.81 & $-$31:47:53.74 & f \\
O43 & ECL & 18.270 & 17.183 & 0.139 & 0.163 & 0.620570 & 7000.37790 & 18:10:35.68 & $-$31:45:26.14 & f \\
O44 & ECL & 16.899 & 15.401 & 0.081 & 0.085 & 12.114004 & 7001.22140 & 18:10:22.82 & $-$31:47:55.70 & f \\
O45 & ECL & 14.563 & 13.903 & 0.044 & 0.050 & 0.635088 & 7000.24380 & 18:10:24.56 & $-$31:45:17.47 & f \\
O46 & ECL & 19.146 & 17.875 & 0.147 & 0.177 & 0.349893 & 7000.22270 & 18:10:29.17 & $-$31:46:54.22 & f \\  
O47 & RRL & 16.731 & 15.526 & 0.357 & 0.250 & 0.672365 & 7000.26497 & 18:10:14.22 & $-$31:45:08.23 & f \\
O48  & DSCT & 17.667 & 16.833 & 0.523 & 0.289 & 0.086767 & 8296.85025 & 18:09:48.88 & $-$31:53:21.90 & ------ \\
O49  & ECL & 16.333 & 15.444 & 0.530 & 0.492 & 0.488999 & 7000.27340 & 18:10:35.97 & $-$31:40:13.90 & f \\
O50  & RRLYR & 15.745 & 14.778 & 0.490 & 0.387 & 0.359998 & 8376.60649 & 18:09:44.80 & $-$31:52:29.00 & ------ \\
O51  & DSCT & 17.114 & 16.292 & 0.446 & 0.270 & 0.121344 & 8334.66273 & 18:10:04.98 & $-$31:51:12.20 & f \\
O52  & ECL & 18.427 & 17.279 & 0.498 & 0.500 & 0.436376 & 7000.36760 & 18:09:52.27 & $-$31:46:30.40 & f \\  
O53 & ECL & 18.222 & 17.084 & 0.351 & 0.374 & 0.491644 & 7000.02930 & 18:10:24.06 & $-$31:41:41.50 & u \\
O54  & ECL & 17.884 & 17.049 & 0.326 & 0.264 & 0.535841 & 7000.09500 & 18:10:31.23 & $-$31:41:26.90 & f  \\
O55  & ECL & 18.004 & 16.996 & 0.442 & 0.299 & 0.681172 & 8296.73116 & 18:10:39.61 & $-$31:41:21.80 & f \\
O56  & DSCT & 17.892 & 16.919 & ------ & ------ & 0.065044 & 8296.63391 & 18:09:44.80 & $-$31:52:29.00 & ------ \\

\hline
\end{tabular}
\end{center}

\quad $*$
m- cluster member, f- field star, u- unknown due to lack of proper motion
\end{table*}

\section{Light curves} 

For all light curves we adopted the same colour code to distinguish from the data source and observing run as follows: green: SWOPE, Blue: BA18, Turquoise; BA19, Black: OGLEIII, Red: OGLEIV, Purple: $Gaia$-DR3.
Fig. \ref{Gaiasources} the light curves of the variables reported by $Gaia$-DR3 are shown. Some stars do not show a clear variability. Clear variations are found in G8 which phases well with a period of 70.985 days, and G10 for which a period of 0.178146 d was found. G8 and G10 were the only stars in this batch  found to be variables, however being field stars we refrained from assigning them a 'V' number.

\begin{figure*}
\includegraphics[width=17.5 cm]{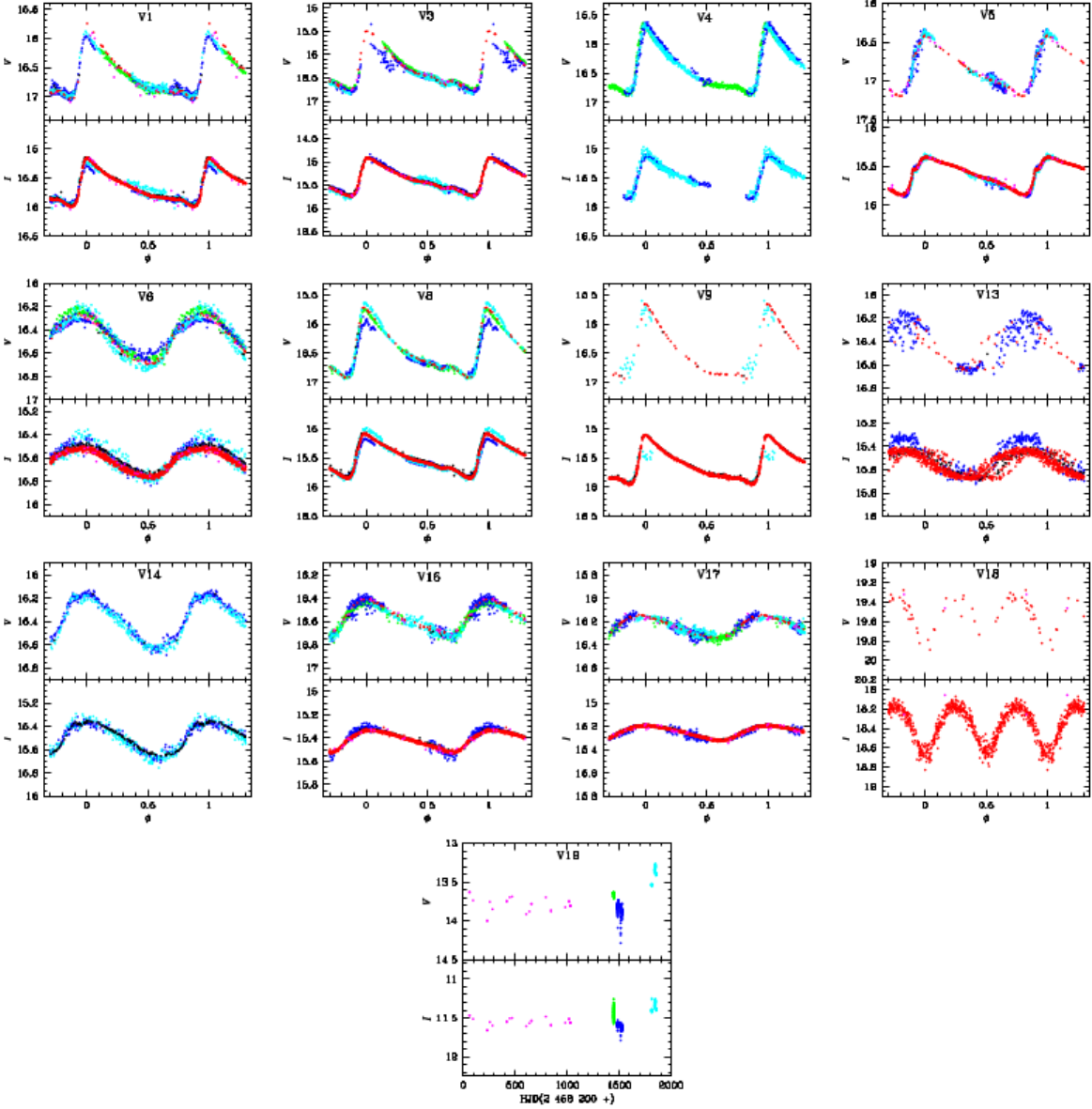}
\caption{Light curves of variables reported the CVSGC \citep{Clement2001} (2016 edition), plus the two new cluster member variables noticed in this work, V18 and V19. Except for V4, where a Blazhko modulation of 26.73 days was found (see section \ref{sec:CMD}), we have not detected secondary frequencies among the scattered light curves. In all cases the scatter is intrinsic to the photometric uncertainties.}
\label{Clem_LC}
\end{figure*}  

\begin{figure*}
\includegraphics[width=17.5 cm]{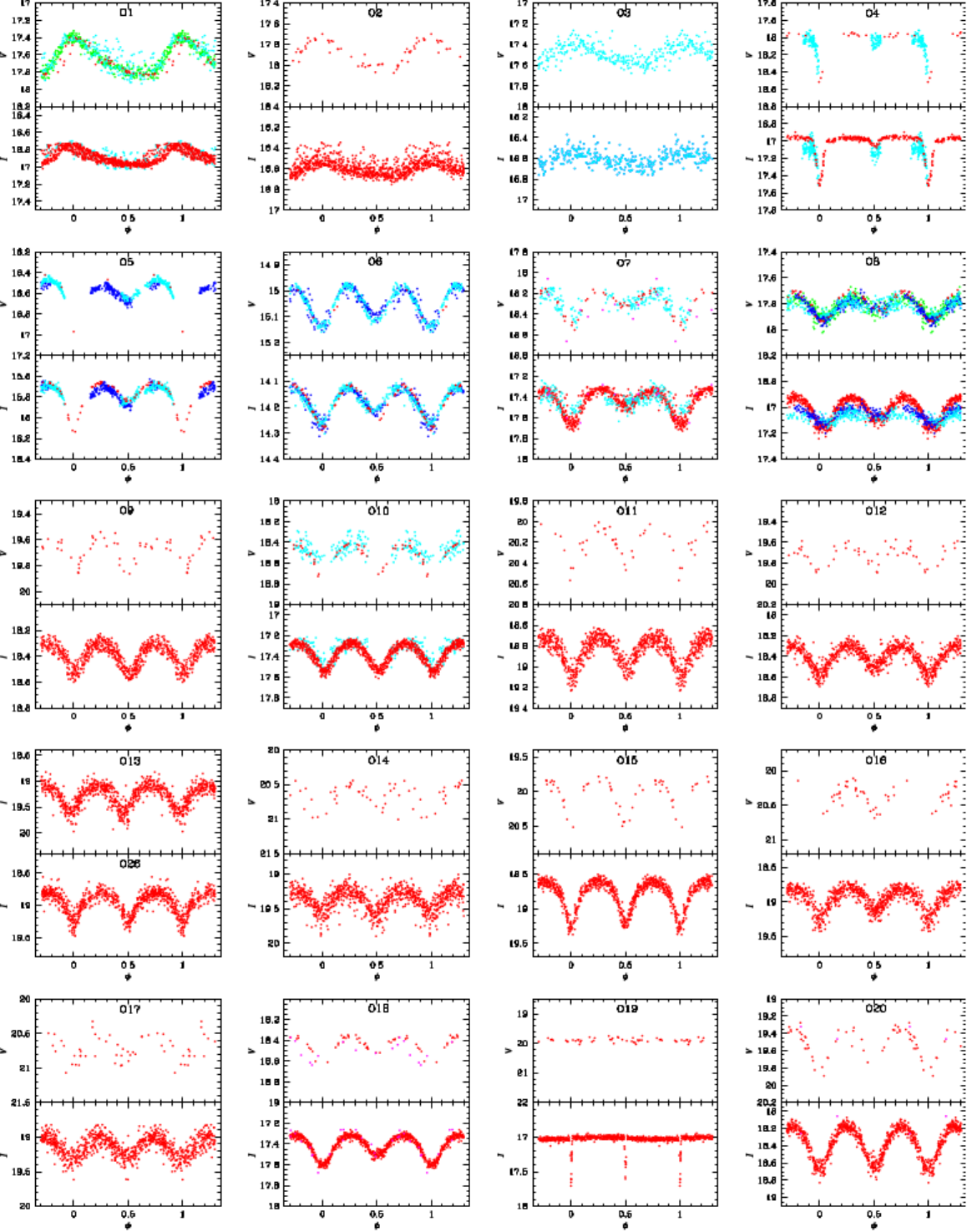}
\caption{Light curves of the 56 variables identified in the OGLE database. A period modulation cannot be ruled out in O50 but our data are insufficient to quantify it.} 
\label{OGLE_vars}
\end{figure*}  

\begin{figure*}
\addtocounter{figure}{-1}
\includegraphics[width=17.5 cm]{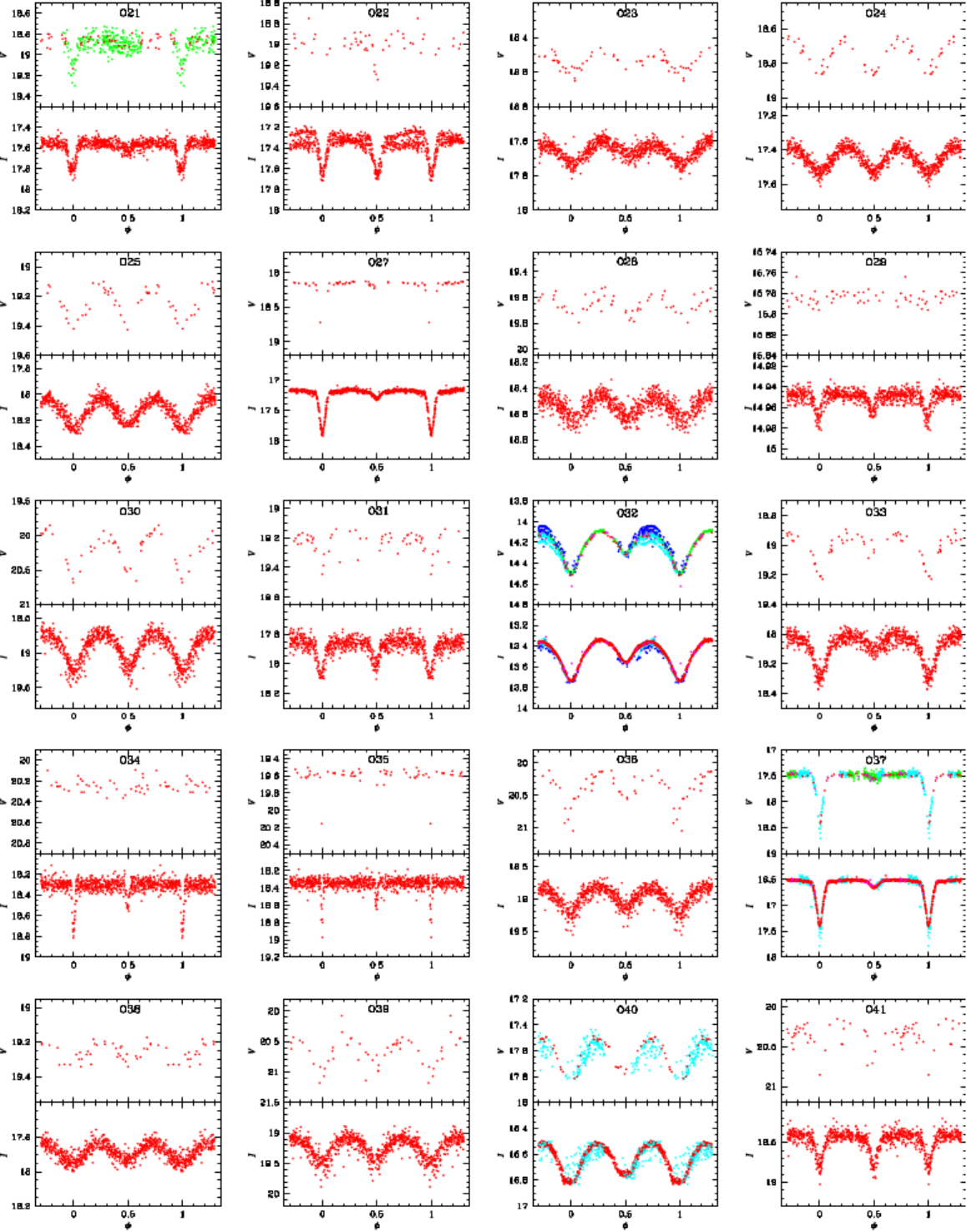}
\caption{Continued} 
\end{figure*}  

\begin{figure*}
\addtocounter{figure}{-1}
\includegraphics[width=17.5 cm]{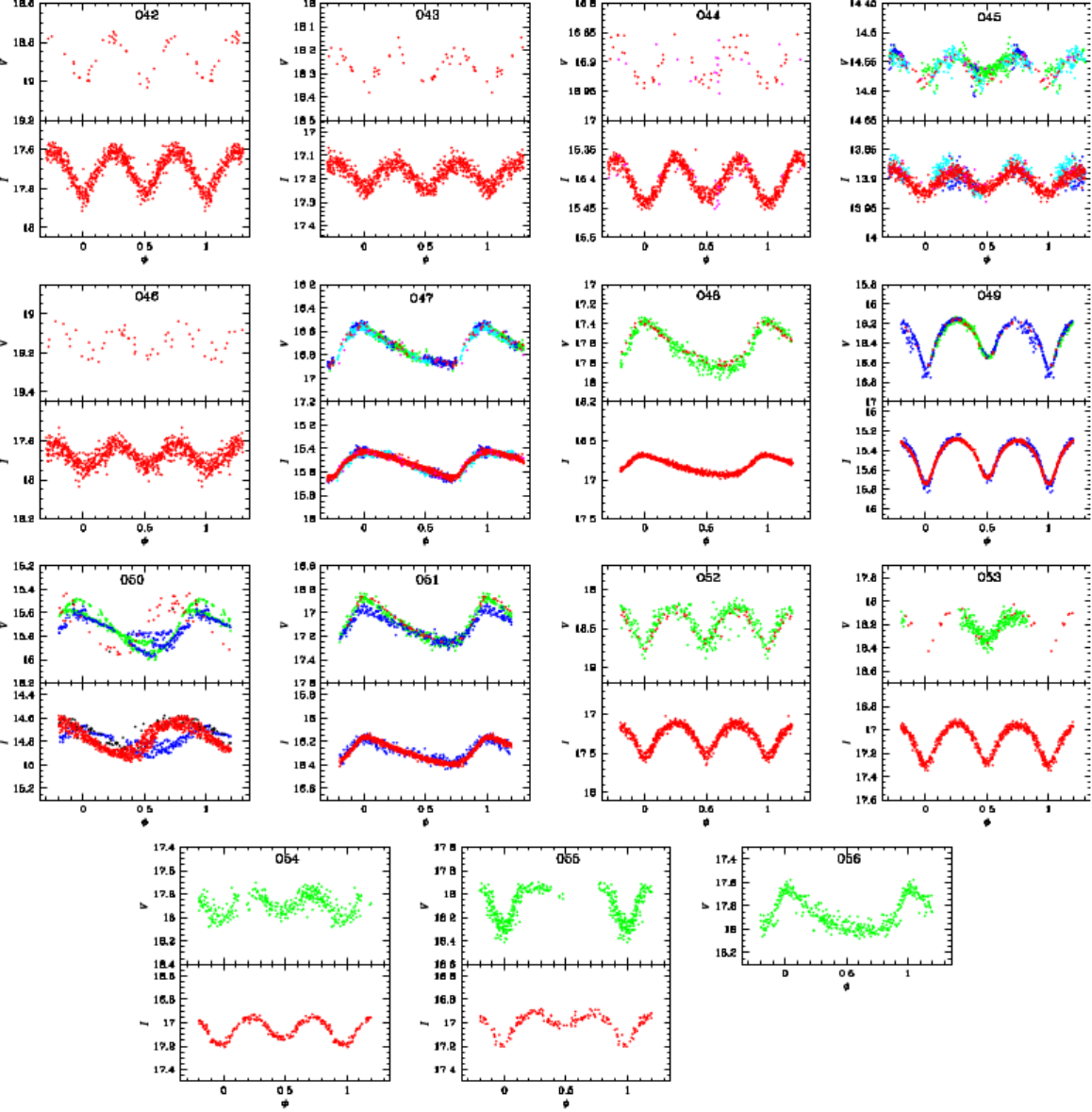}
\caption{Continue.}
\end{figure*}  

\begin{figure*}
\includegraphics[width=16.0cm]{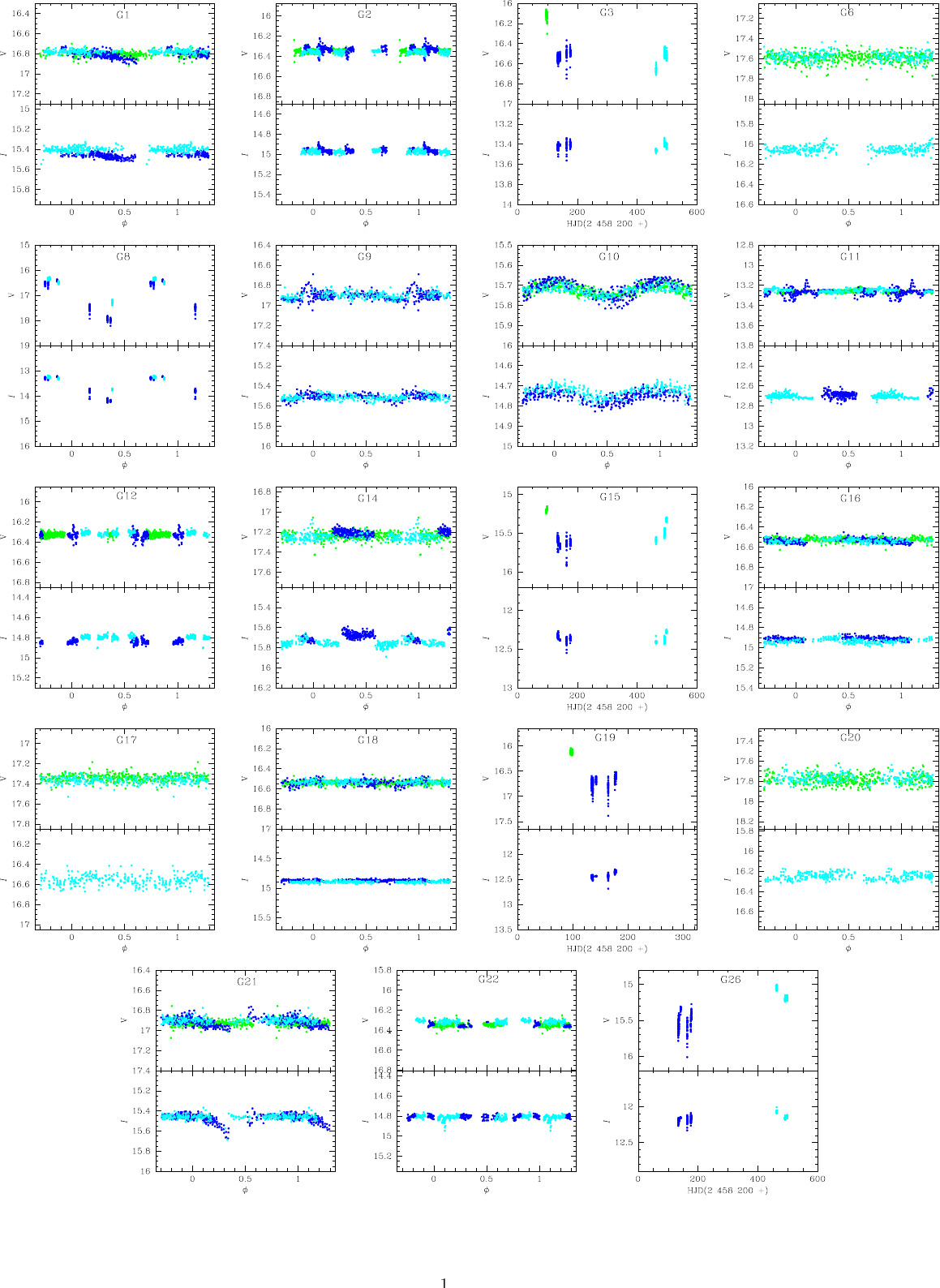}
\caption{Light curves of variables reported in $Gaia$-DR3. Although long-period variability cannot be ruled out when plotted vs HJD in stars G3, G15, G19 and G26, further data would be required to establish them as true variables. Clear variations are found in G8 which phases nicely with a period of 70.985 days, and G10 for which a period of 0.178146 d was found. However, from their proper motion analysis they seem to be field stars.} 
\label{Gaiasources}
\end{figure*}  

\bsp	
\label{lastpage}

\end{document}